# Formation and Dynamics of Vortex Structures in Pure and Gas-Discharge Nonneutral Collisionless Electron Plasmas

## N. A. Kervalishvili

*Andronikashvili Institute of Physics, Javakhishvili Tbilisi State University, Tbilisi 0177, Georgia.   <n_kerv@yahoo.com>*

**Abstract.** The comparative analysis of the results of experimental investigations of the processes of formation, interaction and dynamics of vortex structures in pure electron and gas-discharge electron nonneutral plasmas taking place for the period of time much less than the electron-neutral collision time has been given. The general processes of formation and behavior of vortex structures in these two plasmas were considered. The phenomena, taking place only in one of these plasmas were also considered. It is shown that the existing difference in behavior of vortex structures is caused by different initial states of nonneutral electron plasmas. The role of vortex structures in the processes taking place in nonneutral electron plasma is discussed.

## I. INTRODUCTION

Nonneutral plasmas differ from all known states of any matter. They are the medium consisting of charged particles of only one sign. Therefore, the nonneutral plasmas have their own high electrostatic fields having the strong influence on their behavior and stability. One more peculiarity of nonneutral plasmas that differs them from the usual neutral plasma is the absolute absence of recombination. Therefore, nonneutral plasmas can exist at any temperatures down to absolute zero. Like liquids or gases the nonneutral plasmas consist of the particles of one type, however in nonneutral plasmas the long-range electrostatic repulsive forces are dominant. Under the laboratory conditions, these plasmas can be confined only by strong magnetic fields.

Nonneutral plasma can be of electron, ion or positron type. The electron nonneutral plasma can be conditionally divided into gas-discharge nonneutral electron plasma and pure electron nonneutral plasma, differing from each other by the methods of their obtaining, by the duration of their existence, and by the experimental methods of their investigation.

The gas-discharge nonneutral electron plasma is formed at the expense of ionization of neutral gas by electrons in gas-discharge devices with the crossed electric and magnetic fields of magnetron, inverted magnetron and Penning cell types. The parameters of discharge are such that the ions are not magnetized and leave the discharge gap rapidly without collisions. At the same time, the electrons are strongly magnetized and are trapped by magnetic field. Along the magnetic field, the electrons are hold by electrostatic fields. As a result, near the anode surface, the cylindrical sheath of nonneutral electron plasma is formed, and the whole discharge voltage falls on it. In gas-discharge nonneutral electron plasma there always exist the solitary vortex structures [1-5], that play the main role in the processes of electron transport and dynamic equilibrium of electron plasma sheath. One more peculiarity of gas-discharge nonneutral electron plasma is the ejection of electrons from the vortex structures and from the electron background surrounding them to the end cathodes along the magnetic field. The flux of electrons to the end cathodes always exists in the discharge and the average value of the current of these electrons reaches 50% of the value of discharge current [6,7].

In contrast to gas-discharge plasma, the pure electron plasma is formed by injection of electrons into trap with the crossed electric and magnetic fields. In most investigations, the Penning-Malmberg cell is used as a trap [8-15]. The structure of this cell is very close to Penning cell. Both of these cells is a hollow cylindrical anode located in the longitudinal magnetic field and bounded at the ends by cathodes: flat as in the Penning cell, or in the form of short cylinders as in Penning-Malmberg cell. However, this difference in the form of cathodes is of rather high importance. In the Penning cell, the flat cathodes serve as a source of primary electrons at the expense of ion-electron emission. Therefore, the Penning cell, like a magnetron and an inverted magnetron, is used for ignition of discharge in the crossed electric and magnetic fields. In the Penning-Malmberg cell the primary electrons disappear on the cathodes under the action of the magnetic field, and the ignition of discharge in such geometry is very difficult or absolutely impossible. On the other hand, the Penning-Malmberg cell is ideally adapted for external injection of electrons and for their "extraction" from the cell after passing a certain time. Pure electron plasma in the Penning-Malmberg cell "decays" gradually, as the reproduction of electrons does not take place in it. Therefore, the vortex structures in such plasma are formed only under certain initial conditions or are formed artificially. However, after being formed, they are kept for a rather long time, more than the electron-neutral collision time.

In both electron plasmas the vortex structures are formed with the excess electron density. As the plasma is nonneutral, the excess density means that the vortex structure has its own electric field, and in the presence of external magnetic field rotates around its own axis together with rotation around the axis of experimental device. The vortex structures were investigated experimentally both, in gas-discharge electron and in pure electron nonneutral plasmas. The aim of the present work is the comparative analysis of the processes of formation, dynamics and interaction of vortex structures in these two plasmas, the determination of the general mechanisms, and the explanation of the observed differences.

Before we start the analysis of the properties and behavior of vortex structures in gas-discharge electron and pure electron nonneutral plasmas, let us stop on experimental methods of their investigation. We are not going to consider the all used experimental methods we only mention two main methods of the investigation of vortex structures in these two plasmas. These methods are: the method of two wall probes in the gas-discharge electron plasma [1] and the method of phosphor screen diagnostic in pure electron plasma [8-15].

The method of two wall probes consists in simultaneous measurements of signals from anode and cathode wall probes during the motion of vortex structures around the axis of discharge device. It allows to observe continuously the trajectory and the charge of one or several vortex structures for a long period of time. In combination with the measurement of electron ejection from vortex structures [4], this method gives the possibility to determine the parameters of vortex structures, to study their formation, interaction and dynamics [1-4]. Generally speaking, this method was developed for the geometries of magnetron and inverted magnetron having internal and external cylindrical electrodes. However, its modification is applicable to the case of Penning cell [2].

The method of phosphor screen diagnostic consists in instantaneous ejection of all electrons from the trap to the phosphor screen along the magnetic field. (Sometimes, instead of phosphor screen a movable collector of electrons is used.) This method gives an instant spatial, in ($r,\theta$) plane, pattern of the arrangement and the shapes of vortex structures in the given time moment. However, it is connected with the destruction of plasma imposing strict conditions on repeatability of the process, as each time we investigate the other structures. In order to obtain the exact pattern of the development of process in time, it is necessary to provide for each cycle not only the similar initial plasma parameters but, as well, the formation of vortex structures in one and the same place, at one and the same time, and at one and the same mode of diocotron instability. For this purpose it is necessary to impose the controlled disturbance of electron plasma. Nevertheless, even when these conditions are satisfied, some statistical straggling in the course of the processes of formation, interaction and dynamics of vortex structures still remains.

Thus, the method of phosphor screen diagnostic allows to obtain the full spatial, but statistically averaged pattern of the behaviour of vortex structures, and the method of two wall probes, on the contrary, shows not completely spatial but quite exact and continuous evolution of vortex structures in electron nonneutral plasma. In general, both methods allow to investigate rather in detail the processes of evolution and dynamics of vortex structures. This gives the possibility to make the comparison of these processes for both plasmas. On the other hand, the experimental data obtained by these methods supplement each other making it possible to have a more full pattern of the behaviour of vortex structures in electron nonneutral plasma.

The evolution and the dynamics of vortex structures in gas-discharge and pure electron nonneutral plasmas we consider in two different time intervals: (i) fast collisionless processes ($\Delta t \ll \nu_0^{-1}$) and (ii) slow processes taking place with the participation of electron-neutral collisions ($\Delta t \gg \nu_0^{-1}$). Here, $\nu_0$ is the frequency of electron-neutral collisions. In this paper, the fast collisionless processes are considered. The paper contains both the original oscillograms and illustrations from other works. In section 2 the process of formation of one stable vortex structure in both plasmas is studied. In section 3 the radial displacement of vortex structures in each of plasmas is analyzed. In section 4 the process of approaching and merging of vortex structures is investigated. In sections 5 and 6 the phenomena are considered, that have place only in one of these plasmas: "vortex crystal" in pure electron plasma and "orbital instability" in gas-discharge electron plasma. In final section 7 some experimental results obtained in non-standard devices and role of vortex structures in the processes taking place in nonneutral electron plasma are discussed.

## II. FORMATION OF STABLE VORTEX STRUCTURE

Let us consider the initial time interval $\Delta t \ll \nu_0^{-1}$, during which the processes of formation and evolution of vortex structures take place. In this collisionless time interval we can neglect the ionization, and, therefore, the principle difference between the gas-discharge and pure electron plasmas disappears. Here, all processes should depend only on the initial parameters of electron plasma, and the behavior of vortex structures should be the same for both plasmas. As a starting time interval let us take the moment of appearing the diocotron instability. The diocotron instability in all experiments is a generator of vortex structures with the exception of model experiments in pure electron plasma when the vortex structures are formed artificially.

For appearing the diocotron instability, the nonneutral electron plasma should have the shape of annular sheath (hollow column). Besides, the certain conditions should be fulfilled to excite one or another mode of diocotron instability [16-18].

In gas-discharge electron plasma the annular electron sheath is formed as a result of ionization of neutral gas by electrons. The sheath adjoins the anode surface, but between them there is always a gap of the order of electron Larmor radius. After discharge ignition, the electron density in the sheath increases until it reaches the critical value at which the diocotron instability is excited [17,18]. The minimum critical density corresponds to $l = 1$ mode. The higher is the mode, the more is the critical density of electrons. Therefore, in gas-discharge electron plasma the diocotron instability should be excited at $l = 1$ mode. The experimentally observed instability really has $l = 1$ mode (sometimes $l = 2$ mode) not only in magnetron and inverted magnetron, but as well in Penning cell in which, according to the theory, the exponentially increasing instability at $l = 1$ mode should not be excited.

The annular sheath of pure electron plasma in Penning-Malmberg cell is formed by different ways. In one of them, first, the central electron column is injected with a small seeding asymmetry at $l = 1$ mode; then, a partially hollow profile is formed by means of decreasing the applied potential [8]. By other way, from the very beginning a symmetrical partially hollow profile of electron column is formed. Such electron sheath can be unstable simultaneously for several modes of diocotron instability. For choosing a mode (usually $l = 1$ or $l = 2$) and for a good repeatability of initial conditions of experiment, a controlled initial disturbance of electron plasma is used [8]. One more way consists in the direct injection of annular electron sheath. In this case, as a source of electrons, more frequently, a photocathode is used. Here, from the very beginning, the radius and the thickness of electron ring are given. If a ring is thin the mode of diocotron instability depends on the ratio of the length of the circumference of ring to its thickness, and it can be rather great [12].

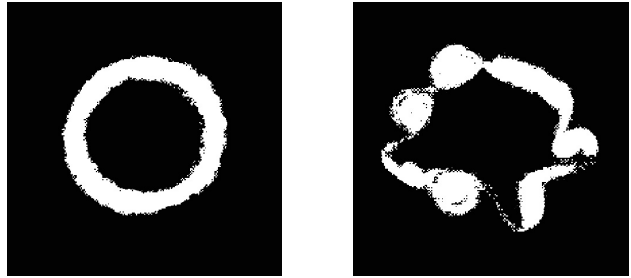

Fig.1. Hollow electron beam [19]
Distance from cathode is 1cm (left) and 8.7cm (right)

Before passing to the consideration of the processes of formation and evolution of vortex structures in both plasmas, let us note that the decay of the annular electron sheath into separate vortex structures was first observed in hollow electron beam that propagated in longitudinal magnetic field [19], and for observation of its profile, a phosphor screen located at different distances from the electron gun was used. It was shown that, beginning from a certain distance, the hollow electron beam decays into separate vortex-like current structures and besides, the neighboring vortex structures are connected with each other by thin spiral "arms". Fig.1 shows the images of electron beam at different distances from the electron gun (photos are taken from [19]).

In gas-discharge electron plasma, it is more convenient to study the processes of formation and evolution of vortex structures in the geometry of inverted magnetron. In this geometry, at low pressures of neutral gas, a full cycle of the evolution of vortex structures can be observed continuously: (i) a fast collisionless process from the appearance of diocotron instability to the formation of one stable (quasistable) vortex structure, and (ii) a slow collisional process of "decaying" the quasistable vortex structure [3,4]. When the vortex structure fully disappears, during a certain period of time before the next diocotron instability appears, only a symmetrical electron sheath remains without vortices and oscillations. The oscillograms of this periodically repeated process are shown in Fig.2. The upper oscillogram is the oscillation of electric field on the anode wall probe. The lower oscillogram is a total electron current on the end cathodes. Here and below, the little lines on the oscillograms (to the left) indicate the initial position of the sweep trace.

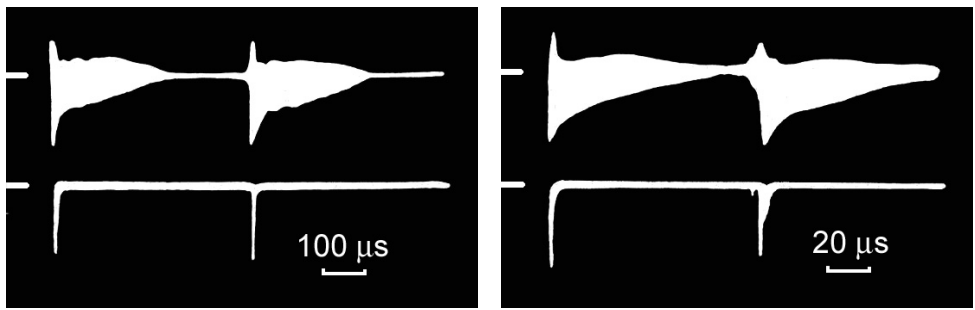

Fig.2. Diocotron instability and vortex structures in inverted magnetron [4]
$r_a = 1.0 cm$; $r_c = 3.2 cm$; $L = 7 cm$; $B = 1.8 kG$; $V = 0.9 kV$; $p = 2\times 10^{-6}$, $1\times 10^{-5} Torr$.

As is seen from the figure, the processes of development of diocotron instability and formation of quasistable vortex structure are accompanied by the pulse of electron ejection from the electron sheath to the end cathodes.

Fig.3 shows the oscillograms of the fragments of this process obtained simultaneously by several oscillographs. The upper oscillograms are the oscillations of electric field on the anode wall probe, and the lower ones – on the cathode wall probe.

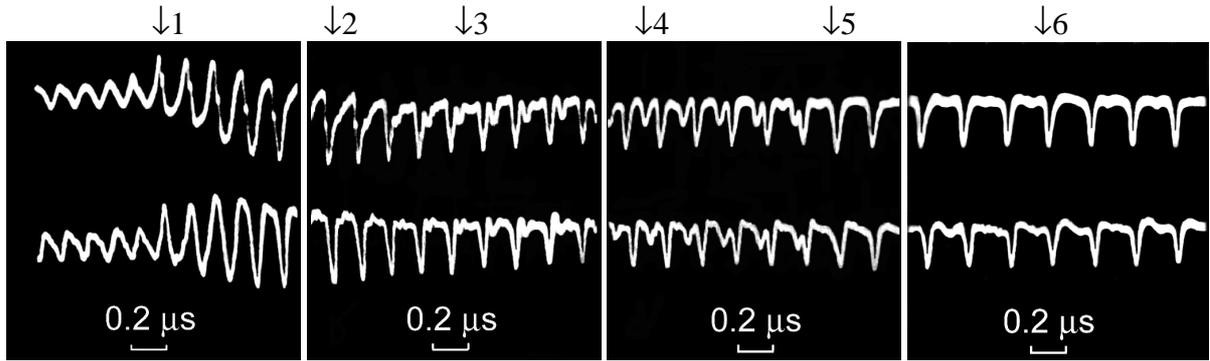

Fig.3. Formation of solitary vortex structure in gas-discharge electron plasma
$r_a = 2.0 cm$; $r_c = 3.2 cm$; $L = 7 cm$; $B = 1.5 kG$; $V = 1.0 kV$; $p = 2 \times 10^{-5} Torr$

Let us consider in detail these oscillograms. The comparison of oscillation amplitudes on the anode and on the cathode gives evidence that the whole process takes place on one and the same drift orbit, and the radial oscillations of vortex structures are absent. In the first part of oscillograms the strongly nonlinear oscillations of diocotron instability are seen. Then, the oscillations increase sharply and in the disturbed region of the sheath a hole (↓1) is formed. The arrow on the oscillogram denotes the place, where the described process takes place. Further, the hole widens along the azimuth, or, speaking strictly, the electrons are bunched at the point diametrically opposite to the azimuth and form a clump of electrons (↓2) followed by a tail. In the tail a lot of small irregular inhomogeneities are formed (↓3). They approach each other, merge and form one more clump (↓4), being smaller than the main one and it moves with less angular velocity. The main clump overtakes the second clump and absorbs it (↓5). At this point, the formation of a single stable vortex structure (↓6) is completed. The formed single vortex structure continues to exist during a long period of time, much longer than the electron-neutral collision time.

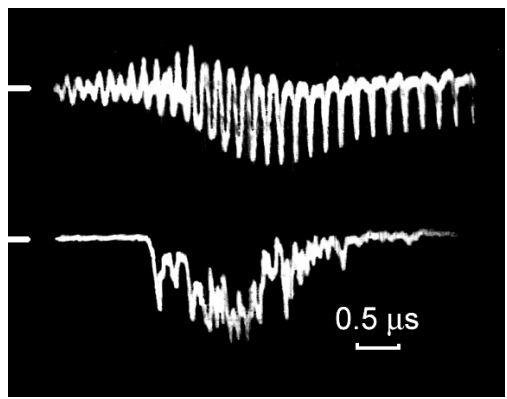

Fig.4. Vortex structure formation and ejection of electrons in inverted magnetron [3]
$r_a = 2.0 cm$; $r_c = 3.2 cm$; $L = 7 cm$; $B = 1.5 kG$; $V = 1.0 kV$; $p = 1 \times 10^{-5} Torr$

It should be noted that the process of formation of a stable vortex structure beginning from the moment of formation of a hole up to the moment of merging the vortex structures (↓1 – ↓5), is accompanied by ejection of electrons along the magnetic field to the end cathodes. This process is shown in more detail in Fig.4, where, like Fig.2, the upper oscillogram is the oscillations of electric field on the anode wall probe, and the lower oscillogram – a total electron current on the end cathodes.

Ejection of electrons take place from the vortex structure and from the electron background (sheath) surrounding it. As the duration of the process of vortex structure formation is less than the time of electron-neutral collisions ($\Delta t \ll v_0^{-1}$), and hence, than the time of ionization, ejection of electrons from the sheath causes the decrease of the average electron density. This should lead to the decrease of the strength of electric field in the sheath and also on the drift orbit of vortex structure. The decrease of electric field on the drift orbit will increase the time of rotation of vortex structure about the axis of discharge device. Indeed, as it follows from Fig.3, the time of rotation of vortex structure increases from 0.15µs for diocotron instability (on the left) to 0.25 µs for the formed stable vortex structure (on the right).

In pure electron plasma, the mode, at which the diocotron instability arises, depends on the initial parameters of electron annular sheath and on the external controlled disturbance of electron plasma. Instability breaks the electron ring into discrete vortex structures, the number of which is equal to the mode of diocotron instability. The vortex structures are mixed, merged (some of them decay) and besides, displaced to the trap axis [8-13]. Finally, one stable vortex structure is left, that is located on the axis of Penning-Malmberg trap. The vortex structures are always followed by filamentary tails that gradually extend, widen, and finally form the symmetric electron background, together with the decayed vortex structures. The density of this background is much less than the density of central vortex structure.

As for $l=1$ mode, it behaves similar to other modes, although, according to the theory, the exponentially unstable $l=1$ mode should not be excited in the geometry with the cathode radius equal to zero. Here, we will consider in more detail the process of formation of vortex structure just at $l=1$ mode, in order to make the comparison with the gas-discharge plasma. In Fig.5 taken from [8] the process of evolution of partially hollow electron column with a small "seeding" asymmetry is shown. Black color of the figure corresponds to the maximum of electron density.

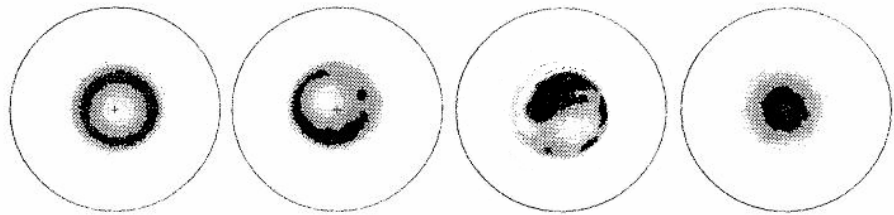

Fig.5. Formation of solitary vortex structure in pure electron plasma [8]
$t = 30, 90, 150, and 1000 \mu s$

As it is seen from the figure, first, the electron ring undergoes breaking. Then, its compression pinching takes place along the azimuth forming a dense clump, i.e. a vortex structure. The vortex structure is followed by a tail (or a arm, as it is called sometimes). On the tail some small clumps are seen, which either disappear or are merged with the main vortex structure. At last, a stable vortex structure is formed, which gradually shifts to the trap axis. At the same time, the central part of low density, on the contrary, displaces outside and diffuses along the azimuth.

Thus, the whole process of evolution from the origination of diocotron instability at $l=1$ mode to the formation of the stable vortex structure in gas-discharge and pure electron plasmas takes place, practically, in a similar way. However, there is one significant difference: in pure electron plasma the stable vortex structure shifts to the trap axis, while in gas-discharge electron plasma it remains near the anode surface.

## III. RADIAL DRIFT OF VORTEX STRUCTURES

Such, ex facte diametrically opposite behavior of vortex structures in pure electron and gas-discharge electron nonneutral plasmas can be explained on the basis of the results of experimental and theoretical investigations carried out in pure electron plasma. In [13] the evolution of a thin annular electron sheath was investigated depending on the value of its radius. An annular electron sheath was formed by a photocathode, and the result of its evolution was observed on the phosphor screen. Fig.6 presents the photos taken from [13] that show the evolution of two thin electron rings of different radii during one and the same time interval. In both cases, as a result of diocotron instability, the ring breaks into discrete vortex structures. After a small period of mixing and merging, only one stable vortex structure and symmetrical electron background of lower density are left. If the average radius of annular sheath $r_i$ is less or of the order of $0.5\, r_a$, where $r_a$ is the anode radius, the vortex structure shifts to the center, and the background falls from the center of the trap to the wall. But if $r_i$ is more than $0.5\, r_a$, the stable vortex structure remains shifted from the trap axis, and the background reaches the wall of the trap and falls towards its center.

Now it is clear why there is a difference in the behavior of stable vortex structure in gas-discharge and pure electron plasmas. In gas-discharge electron plasma the electron sheath adjoins the anode surface. Therefore, the formed solitary vortex structure remains as well near the anode surface. The pure electron plasma is injected, as a rule, to the central part of Penning-Malmberg trap. Therefore, the solitary vortex structure formed there is shifted to the trap axis.

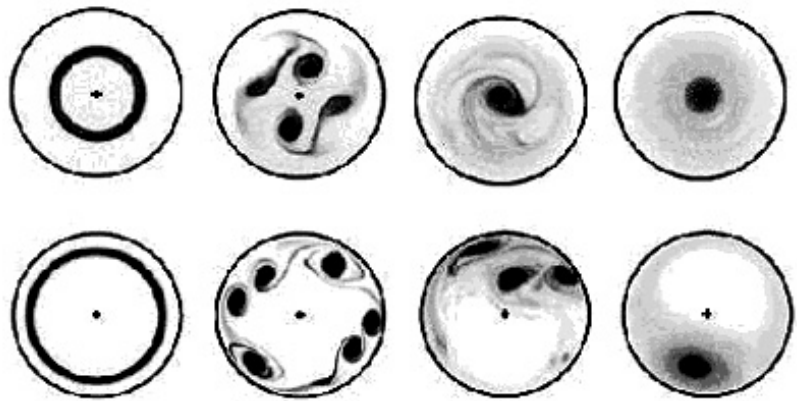

Fig.6. Time evolution of an electron rings [13]
Top: $(r_i/r_a) = 0.5$, $t = 0.01, 3, 10, 30\, ms$. Bottom: $(r_i/r_a) = 0.79$, $t = 0.01, 1, 5, 30\, ms$.

The mechanism of radial drift of vortex structures to the axis of Penning-Malmberg trap in pure electron plasma was investigated theoretically in [20, 21], where the motion of point vortex structure is considered in axisymmetric nonuniform electron background with the shear of velocities. The maximum density of electron background is on the trap axis. The vortex having the density more than the background density is a "clump", and the vortex having the density less than the background density is a "hole". The authors define both, the clumps and the holes as prograde or retrograde depending on the fact whether they rotate with or against the local shear. According to such definition, for the background with the maximum on the trap axis the clumps are retrograde, and the holes – prograde. The work shows that under the condition of conservation of canonical angular momentum, a clump should move up in the background gradient, i.e. to the trap axis, and a hole – down in the background gradient, i.e., radially outside. Besides, the velocity of a prograde is by an order of magnitude less than that of retrograde. The obtained results [20, 21] correspond well to the experimental data of [10].

Let us see how this theory is qualitatively applicable to gas-discharge electron plasma. In gas-discharge electron plasma in all three geometries of discharge device – in Penning cell, in magnetron and in inverted magnetron – the density of electron background increases towards the anode and has the maximum near the anode surface. Hence, the radial drift of vortex structures "clumps" at the expense of the gradient of electron background in gas-discharge electron plasma will be always directed to the anode. However, the vortex structures cannot approach the anode nearer than the maximum of electron background is located. Therefore, they will be at some distance from the anode surface. In all these cases, according to the definition of [21], the vortex structures "clumps" will be prograde. Therefore, the velocity of their radial drift will be much less than in pure electron plasma, when the clump drifts to the axis of the cell. In the inverted magnetron the electron sheath is paramagnetic and rotates to the direction opposite to the direction of rotation of diamagnetic sheath in the magnetron geometry, while the rotation of vortex structures in both geometries is the same. Hence, in spite of the fact that the density of sheath electrons in the inverted magnetron increases towards the axis of discharge device, the vortex structures "clumps" in the inverted magnetron will be prograde too.

Thus, the theory of radial shift of vortex structures at the expense of the gradient of the electron background density is in qualitative agreement with the experimental results not only in pure electron plasma but also in gas-discharge electron plasma. However, there is one uncertainty regarding the conservation of canonical angular momentum in gas-discharge electron plasma. The formation and the interaction (approach, merging) of vortex structures are accompanied by the pulsed ejection of electrons to the end cathodes along the magnetic field both, from the vortex structures themselves and from the region of electron sheath (background) adjoined to them [1-5]. Under such conditions, it is difficult to say whether the canonical angular momentum is conserved.

It should be noted that the conditions for the ejection of electrons from the vortex structures for both nonneutral electron plasmas are not similar. At the formation and at the approach (merging) of vortex structures, as well as at the approach of the vortex structure to the trap axis the potential barrier between the vortex structure and the cathode decreases. If, at the same time, a part of the electrons of vortex structure has such a longitudinal velocity which allows them to overcome this decreased potential barrier, the ejection of these electrons to the end cathodes along the magnetic field will take place. At the consideration of collisionless situation, we should take into account the initial longitudinal electron velocity. For the pure electron plasma, it is determined by the temperature of injected electrons and by the process of filling the trap, and for the gas-discharge electron plasma – by the previous electron-neutral collisions. In general, one can assume that the conditions for the ejection of electrons in gas-discharge electron plasma are preferable.

**IV. INTERACTION OF VORTEX STRUCTURES**

When in nonneutral electron plasma there exists simultaneously several vortex structures, the problem of their interaction becomes important. The interaction of vortex structures in the time interval $\Delta t \ll v_0^{-1}$ was studied in both plasmas.

In the pure electron plasma several stable vortex structures are formed either at the expense of diocotron instability at $l > 1$ mode, or as a result of injection of several electron columns into Penning-Malmberg trap. In the gas-discharge electron plasma at the low pressures of neutral gas, one stable vortex structure is formed as a result of diocotron instability. However, at the pressures of the order or more than $5 \times 10^{-5} Torr$, there exists simultaneously several vortex structures, and higher is the pressure, the more is their number. These vortex structures do not depend on each other. They move on different circular orbits with different angular velocities, and therefore, approach each other periodically. Fig.7 shows the oscillograms depicting the process of approaching the vortex structures.

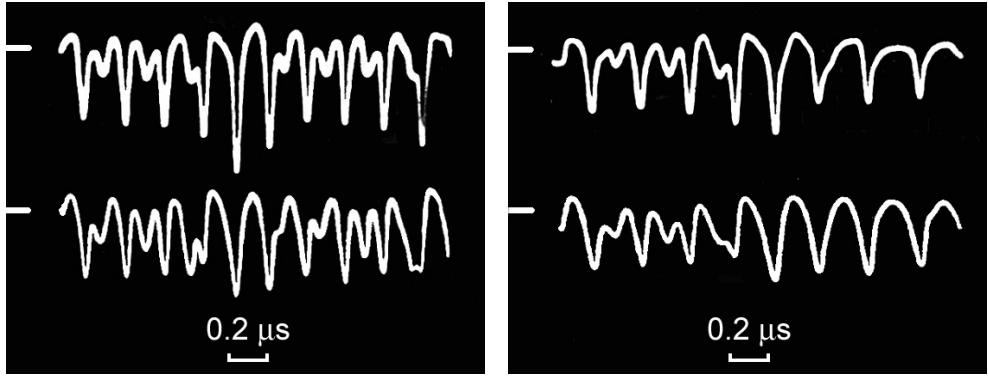

Fig.7. Approach and merging of vortex structures in magnetron
$r_a = 3.2cm$; $r_c = 1.0cm$; $L = 7cm$; $B = 1.5kG$; $V = 1.5kV$; $p = 1\times10^{-4}Torr$.

The upper oscillograms present the oscillations of electric field on the anode wall probe, and the lower - the oscillations of electric field on the cathode wall probe. On the left the process of passing of one vortex structure by the other one is shown, and on the right – the process of merging of two vortex structures. In the gas-discharge electron plasma the process of approaching the vortex structures is accompanied by the pulsed ejection of electrons along the magnetic field both, from the vortex structures themselves and from the adjoining region of electron sheath (background) [1-3]. This is shown in Fig.8, where the upper oscillograms present the oscillations of electric field on the anode wall probe, and the lower - the full current of electrons on the end cathodes. On the left, the oscillograms are given for two vortex structures, and on the right - when the number of vortex structures is more than two.

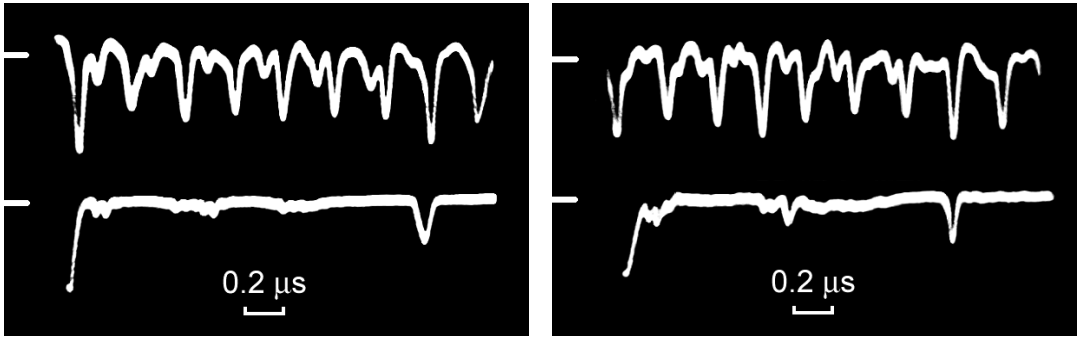

Fig.8. Approach of vortex structures in magnetron
$r_a = 3.2cm$; $r_c = 1.0cm$; $L = 7cm$; $B = 1.5kG$; $V = 1.5kV$; $p = 1\times10^{-4}Torr$.

As the vortex structures move on different drift orbits, the probability of their merging depends on how near they approach each other at passing one vortex structure by the other one. The process of merging of vortex structures and the conditions under which this takes place, were studied in detail by the experiments in pure electron plasma [9]. In these experiments, two electron columns of the given diameter and with the given distance between them were formed artificially, and then the process of their approach and merging was studied.

Fig.9 taken from [9] shows the contours of electron densities in $(r,\theta)$ plane measured in the process of merging of two equal vortex structures.

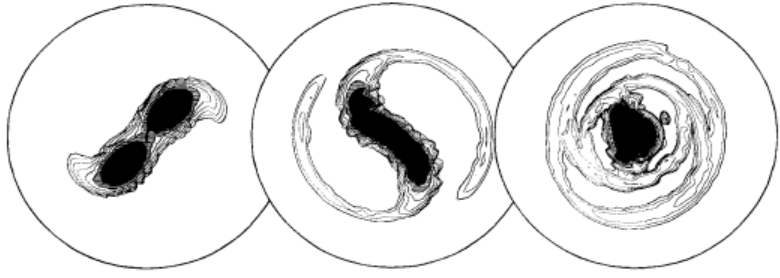

Fig.9. Merging of vortex structures [9].
$D/2R_v = 1.48$, $t = 10, 40, and\ 70\mu\sec$

As is seen from the figure, the vortex structures are followed by the filamentary tails that are mixed and form the background of low density. It appeared that the time of merging of vortex structures depends critically on the ratio of the distance between their centers to their diameter $(D/2R_v)$. If $D/2R_v < 1.6$, the merging of vortex structures takes place for several orbital time. If $D/2R_v > 1.7$, two vortex structures make more than $10^4$ rotations about each other until they merge. These results are in good agreement with numerical simulations and two-dimensional ideal fluid theory.

**V. "VORTEX CRYSTALS" IN PURE ELECTRON PLASMA**

Above we investigated the general mechanisms in the behavior of vortex structures in gas-discharge and pure electron plasmas. However, there are the phenomena observed only in one of these plasmas. First of all, this is the "vortex crystals" in pure electron plasma and the "orbital instability" in gas-discharge electron plasma.
The vortex crystals are rigidly rotating equilibrium lattices of intense vortices of small diameters in the background of lower vorticity. Such vortex crystals are formed, e.g. in the rotating vessel with superfluid helium [22], at the same time, the number of vortices in the lattice increases with the increase of the velocity of vessel rotation.

In [11] it was shown that the vortex crystals can be formed in pure electron plasma. This process is depicted in Fig.10 taken from [11]. In this experiment the layered electron column being a thin flat electron sheath reeled in a roll was injected into the trap. The strong magnetic field inhibited the mixing of layers. As a result of local diocotron instabilities, a great number of individual vortex structures (clumps) were formed. The turbulent state appearing in this case was evolved and relaxed at the expense of chaotic mixing and merging of vortex structures. A part of vortex structures decayed forming the electron background. The final state, generally, was the solitary vortex structure located at the center with approximately initial density of electrons and the electron background of low density surrounding it. The whole process took place during about 10 rotation times $\tau_R$ (the lower sequence in Fig.10). However, sometimes, the relaxation was stopped and the individual vortex structures were located in the form of regular vortex lattice (the upper sequence in Fig. 10). Such quasistationary state lasted about $10^4 \tau_R$, after which the number of the vortex structures decreased to one located at a center of the device. The decrease of the number of vortex structures took place step-wise. After each decrease the remained vortex structures rearranged into a new rigidly rotating symmetric quasistationary configuration.

The experimentally found [11] crystal lattices of electron vortex structures in pure electron plasma were confirmed by the numerical simulation [23] and reproduced theoretically [24]. In [11, 23, 24] it was shown that the formation of vortex crystal takes place as a result of interaction between the vortex structures (clumps) and the background of low density surrounding them.

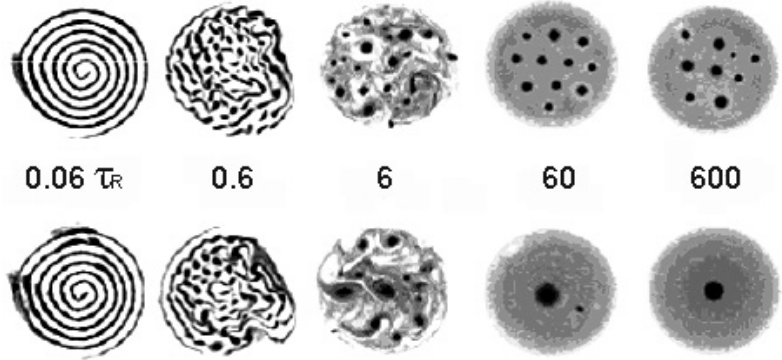

Fig.10. Relaxation of 2D turbulence to vortex crystals and to a single vortex [11]

The systematic experimental investigation of the contribution of electron background to spontaneous formation and decay of vortex crystals was given in [14,15]. In these experiments, the clumps and the background were generated separately and their superimposition formed the initial state of the system. The background of the given value and profile was formed preliminarily. Then, the clumps with the given densities and location were injected in it. This allowed to create the controllable and well-reproducible conditions of experiment. As the experiments showed, the vortex crystals are not formed in vacuum. For formation of vortex crystals, the presence of the background is necessary, and besides, the density of electron background should exceed the definite level necessary for the given number of clumps. Interaction of clumps with the background leads to the formation of annular zones around the clumps depleted of electrons. These annular zones hold the electron vortex structures in lattice points and prevent them to approach each other, even for the unequal charges of vortex structures. In Fig.11 taken from [14] such zones are shown for three vortex structures with different charges.

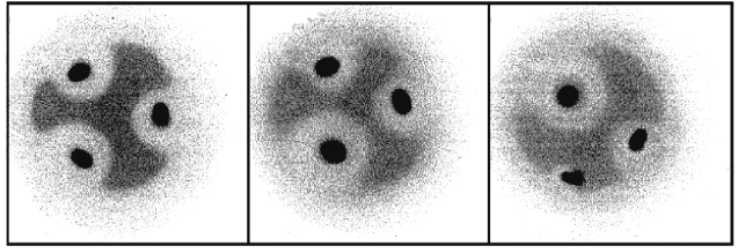

Fig.11. Ring holes around the clumps in vortex crystal [14]

Thus, the vortex crystals are formed, when the dense vortex structures (clumps) are in the electron background of low density and the level of background density has the value necessary for the given number of clumps. If the initial value of background density is not high enough, the number of clumps decreases in the process of free relaxation until the value of background is sufficient for the remained number of clumps to form a crystal. After the crystal is formed, the configuration remains quasistationary until one of the clumps disappears thanks to some dissipative processes. Disappearance of a clump destroys the stable configuration and triggers the turbulent vortex dynamics lasting until a new stable configuration is formed.

**VI. "ORBITAL INSTABILITY" IN GAS-DISCHARGE ELECTRON PLASMA**

In gas-discharge nonneutral electron plasma the different unexpected phenomena take place. At low pressures of neutral gas ($p < 10^{-5} Torr$), in magnetron geometry and in Penning cell a periodically appearing instability of orbital motion of single vortex structure was observed [1,2]. Instability manifests itself in creation of strong radial oscillations of vortex structure the frequency of which is almost by an order of magnitude less than the frequency of its orbital motion. Fig.12 shows the oscillograms of oscillations of electric fields on the anode (the upper) and the cathode (the lower) wall probes during this instability in the magnetron geometry of the discharge device.

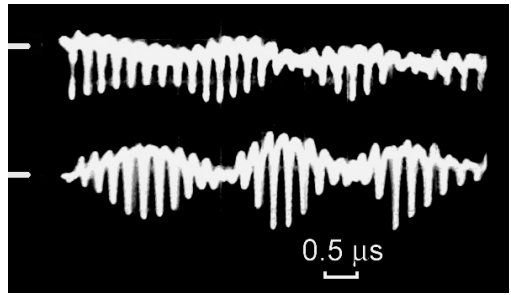

Fig.12. Orbital instability in magnetron
$r_a = 3.2 cm$; $r_c = 1.0 cm$; $L = 7 cm$; $B = 1.2 kG$; $V = 1.5 kV$; $p = 5 \times 10^{-6} Torr$.

As it is seen from the figure, the vortex structure moves away from or approaches the anode periodically. As the period of radial oscillations of vortex structure is much more than the time of its rotation about the axis of the discharge device, the vortex structure makes the motion in spiral. When the vortex structure moves away from the anode, it losses a part of its electrons that escapes along the magnetic field. This is well seen in Fig.13, presenting the oscillograms of oscillations of the electric field on the cathode wall probe (the lower) and of the full electron current on the end cathodes (the upper).

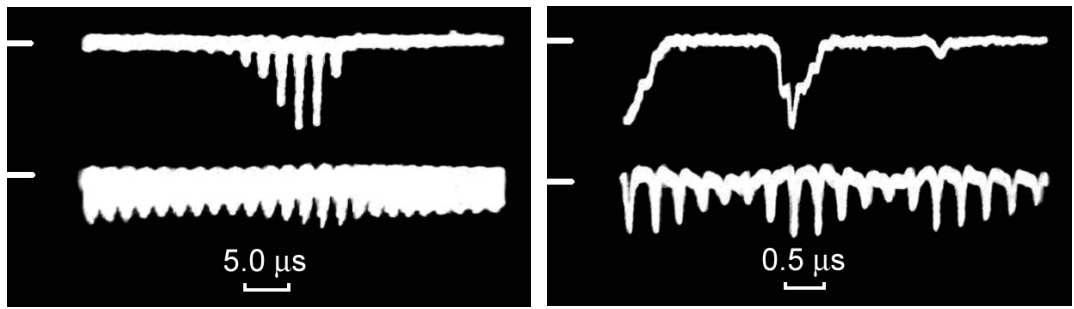

Fig.13. Orbital instability and electron ejection in magnetron
$r_a = 3.2 cm$; $r_c = 1.0 cm$; $L = 7 cm$; $B = 1.2 kG$; $V = 1.5 kV$; $p = 5 \times 10^{-6} Torr$.

The amplitude of radial oscillations of vortex structure is rather large and reaches 1cm [1]. After making 5-8 radial oscillations and losing about one third of its charge, the vortex structure "calms down". The orbit of vortex structure becomes stable. However, now the radius of vortex structure orbit becomes by 5-6 mm less than it was before starting the instability. During the instability, the vortex structure is not destroyed, only its orbital motion becomes unstable. Therefore, such instability we will call "orbital instability". After completing the orbital instability, the charge of vortex structure and the radius of its drift orbit start to increase slowly, but this is already the collisional process, which continued until the orbital instability is repeated again.

The orbital instability in Penning cell takes place in the same way as in magnetron geometry, though there are some differences. The number of radial oscillations in Penning cell is more than in magnetron geometry (of the order of ten and more). The decay of radial oscillations in the Penning cell in most cases, takes place in the same way as in magnetron geometry, that is, the amplitude of radial oscillations and the amplitude of electron ejection along the magnetic field decrease gradually (Fig.14 left). Here the upper oscillogram is the oscillations of electric field on the anode wall probe, and the lower – the current of electrons on the cathodes.

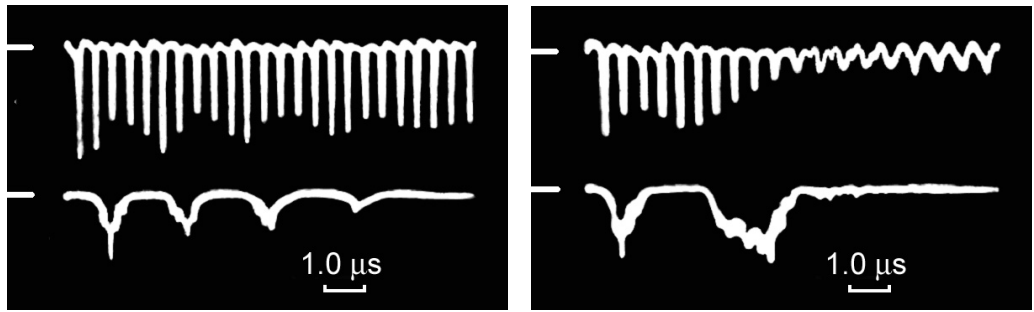

Fig.14. Orbital instability in Penning cell
$r_a = 3.2 cm$; $L = 7 cm$; $B = 2.0 kG$; $V = 2.0 kV$; $p = 6 \times 10^{-6} Torr$

However, in some cases, the last ejection of electrons appears to be the greatest and the amplitude of oscillations on the anode wall probe decreases strongly (Fig.14 right). This indicates that, sometimes, the vortex structure approaches the axis of Penning cell quite close causing, thus, the great losses of electrons in the vortex structure. Then the vortex structure remains near the axis of Penning cell during the whole collisionless period. The mechanism of appearing and progressing of orbital instability is not studied yet.

## VII. CONCLUSION

We considered the processes of formation and dynamics of vortex structures taking place during the collisionless time interval in pure electron and gas-discharge electron nonneutral plasmas. The analysis of experimental results showed that the process of formation of stable vortex structure as a result of diocotron instability at $l = 1$ mode takes place in similar way in both plasmas. If, as a result of diocotron instability several vortex structures are formed, in the process of their evolution and interaction one stable vortex structure is left finally. A great role in both plasmas is played by electron background that determines the direction of radial drift of vortex structure and the conditions of vortex crystal formation.

The difference in behavior of vortex structures in pure electron and gas-discharge electron nonneutral plasmas is caused, mainly, by the different initial conditions. In pure electron plasma there is a wide spectrum of externally given initial conditions: the thickness, the location and the density of circular electron sheath, or the given arrangement of electron columns. Therefore, at the initial moment we can obtain any mode of diocotron instability and any number of vortex structures. However, finally, as a result of mixing and merging, one stable vortex structure is left. If a vortex crystal is formed, in this case as well the number of vortex structures decreases to one in step-wise manner.

In gas-discharge electron plasma the situation is somewhat different. Here, the cylindrical annular electron sheath is formed by natural way at the expense of collisions and ionization. Therefore, the diocotron instability develops at $l = 1$ mode, corresponding to the minimum critical electron density [17]. This leads directly to the formation of one stable vortex structure. Thus, in gas-discharge electron plasma there are not possibilities for variation of initial conditions. in contrast to the case of pure electron plasma. However, here, one can vary the geometry (magnetron or inverted magnetron) of discharge device depending on the fact in which of these geometries we can observe better one or another process.

Generally, the use of different geometries or of their modifications is very useful for carrying out the experiments. In [25] the vortex structures in magnetron geometry with dielectric end plates were studied experimentally. It was found that in such discharge device, the diocotron instability appears periodically and a quasistable vortex structure is formed. The existence of dielectric discs inhibits the ejection of electrons along the magnetic field. In this regard, the situation reminds the one taking place in pure electron plasma. Periodical formation of vortex structure and its subsequent "decay" reminds the situation connecting with gas-discharge plasma in inverted magnetron. In [26] the experimental investigation of the dispersion shifted from the axis of vortex structure in Penning-Malmberg cell was made. For creation of controllable radial electric field, a long thin wire was stretched along the cell axis, to which the displacement potential was applied. This allowed to study the process of evolution and dispersion of vortex structure in an applied irrotational shear flow. In [27] a uniform annular sheath of magnetized electrons was formed between two coaxial cylinders by means of continuous electron injection. Diocotron instability in annular electron sheath led to the formation of stable localized vortex structures of high density. In [28] the electron column was continuously injected into Penning-Malmberg cell. With the increase of electron density, the column became hollow. The diocotron instability appeared and the vortex structures were formed. However, in contrast to the case of short-time injection, when the formed vortex structures shifted to the cell axis, in the given experiment the vortex structures shifted to the region between the electron column and the anode. For simulation and studying the magnetospheric phenomena, the device was created with levitation superconducting dipole magnet for confining the pure electron plasma with magnetospheric configuration [29]. It was shown that in this device the electrons, similar to the experiments described above, are self-organized into dense stable vortex structures.

Thus, the self-organizing vortex structures (clumps) are formed, practically, always in nonneutral electron plasma independent of the method of its obtaining and of the geometry of confining device. Therefore, the vortex structures are one of the fundamental properties of such plasma. It should be noted that nonneutral plasma always has the limited dimensions along the electric field due to the existence of large internal electric fields. Therefore, it can exist either in the form of the sheath with definite thickness and with the shear of velocities that leads to the development of diocotron instability and, hence, to the formation of vortex structures, or in the form of the column rotating around its own axis being, thus, a vortex structure. Thanks to the excess electron density, each vortex structure has its own electric field. Therefore, it interacts both, with the electron background, and with the other vortex structures at sufficiently large distances. Thus, even a single vortex structure, if, especially, it is located off axis of electron background, has a strong influence on the processes taking place in nonneutral electron plasma. The electron sheath (background), in its turn, has the influence on the behavior of vortex structures, for example, on the formation of vortex crystals and on the radial drift of vortex structures. Therefore, nonneutral plasma with vortex structure is the complex dynamically system, at considering of which it is necessary to take into account both of its components.